\documentclass{jfm}
\usepackage{graphicx}
\usepackage{epstopdf, epsfig}

\usepackage{xcolor,soul}
\usepackage{lineno}

\shorttitle{The speed of breaking waves controls sea surface drag}
\shortauthor{A. Ayet, B. Chapron, P. Sutherland and G.G. Katul}

\title{The speed of breaking waves controls sea surface drag}

\author{Alex Ayet\aff{1,2}
  \corresp{\email{alex.ayet@normalesup.org}},
  Bertrand Chapron\aff{1},
  Peter Sutherland\aff{1}
 \and Gabriel G. Katul\aff{3}}

\affiliation{\aff{1}Ifremer, CNRS, IRD, Univ. Brest/ Laboratoire d'Oc\'eanographie Physique et Spatiale (LOPS), IUEM, Brest, France
\aff{2}LMD/IPSL, CNRS, \'Ecole Normale Sup\'erieure, PSL Research University, Paris, France
\aff{3}Nicholas School of the Environment and the Department of Civil and Environmental Engineering, Duke University, Durham, NC 27708-0328, USA}

\begin{document}

\maketitle

\begin{abstract}
The coupling between wind-waves and atmospheric surface layer turbulence sets surface drag. This coupling is however usually represented through a roughness length. Originally suggested on purely dimensional grounds, this roughness length does not directly correspond to a measurable physical quantity of the wind-and-wave system. Here, to go beyond this representation, we formalize ideas underlying the Beaufort scale by quantifying the velocity of breaking short waves that are the most coupled to near-surface wind. This velocity increases with wind speed, reflecting the fact that stronger winds can be visually identified by longer (and faster) breakers becoming predominant on the sea surface. A phenomenological turbulence model further shows that this velocity is associated with breaking waves that impede the most the formation of turbulent eddies. Scales of such eddies are then constrained inside a so-called roughness sub-layer. Unlike previous theoretical developments, the proposed breaker velocity is a directly measurable quantity, which could be used to characterize the coupling between wind and waves using remote sensing techniques. This work provides a physical framework for new formulations of air-sea momentum exchange in which the effects of surface currents and slicks on surface drag can also be incorporated. Finally, it provides a long-sought physical explanation for the Beaufort scale: a universal link between wave breaking, wind speed and surface drag.

\end{abstract}


\section{Introduction}

The interaction between near-surface turbulence (up to a few meters from the surface) and wind-waves (waves shorter than ten meters) determines air-sea fluxes of momentum, heat and gases.  Yet,  this link is not readily expressed in current parameterizations of surface drag, which rely on a roughness length \citep[e.g.][]{kitaigorodskii1973physics, edson2013exchange}. Originally suggested on purely dimensional grounds \citep{charnock1955wind}, this roughness length does not directly correspond to a measurable physical quantity of the wind-and-wave system \citep{kraus1967wind}. More importantly, it does not reflect the fact that the wind blows over a multi-scale and moving surface (each wave moving with its own phase speed). As such, it is difficult to capture (experimentally and in parameterizations) the sensitivity of air-sea fluxes to external parameters such as ocean surface currents, which alter the speed of wind-waves and hence surface drag.

However, seafarers have known for centuries how to estimate wind velocity by observing the local sea surface. A windy sea indeed demonstrates an apparent visual organization associated with the occurrence and intensity of breaking waves. This led George Simpson, in 1906, to derive a scale for the surface wind speed, labeled the Beaufort scale after the original classification of the sea surface by Frances Beaufort in 1831. Such an estimation reflects the existence of physical quantities, related to short-scale breaking waves, that could be used to characterize the wind-and-waves system and hence air-sea fluxes. This work presents one such quantity, aimed at characterizing the momentum flux at the air-water interface.

In fact, it has long been known that the statistics of short breaking waves are tightly linked to the local properties of near-surface turbulence, surface drag, and wind speed, both from measurements \citep{melville1985momentum, katsaros1992dependence} and from theoretical studies \citep{melville1977wind, phillips1985spectral, newell1992rough, Kudryavtsev2014}. 
On the one hand, breaking waves are steep disturbances of the sea-surface that influence its overall roughness \citep{melville1977wind, Kudryavtsev2014}. On the other hand, wave breaking and the formation of whitecaps play a fundamental role in the energy transfer between wave scales, and hence in the slope distribution of short waves \citep{phillips1985spectral}, which is ultimately related to surface roughness and drag \citep{munk1955wind}. Including this link in parameterizations is difficult due to the challenges associated with modeling a surface that is no longer connected (i.e. disrupted by wave breaking), and for which the governing equations of motion are not fully known \citep{newell1992rough}.

    
Recently, technological advances resulted in high resolution images of the sea surface, which can now be used to advance our understanding of short-scale wave breaking statistics \citep{sutherland2013field, sutherland2015field}. In parallel, phenomenological models of turbulence close to a boundary have emerged \citep{gioia2010spectral}: those models establish links between the spectral properties and mean properties of the turbulent flow \citep{katul2014cospectral, ayet2020scaling}, and have been used to describe their modification due to changes in atmospheric stability \citep{katul2011mean}, the presence of roughness elements \citep{gioia2001scaling, bonetti2017manning} and, recently, the presence of long wind-waves \citep{ayet2020impact}. 

In this work, both advances are used to show that the complex interaction between short, breaking wind-waves and wind can be expressed as function of a representative velocity of short-scale breaking fronts. The wave properties associated with the predicted phase speed correspond to the Beaufort scale. This velocity is derived independently (i) from high resolution measurements (in \S\ref{sec:data}) and (ii) by extending a phenomenological model of turbulence (in \S\ref{sec:pheno}). Finally, in \S\ref{sec:tri} we show that this representative speed is associated with breaking waves that efficiently impede the generation of turbulent eddies. Unlike previous theoretical developments, this velocity is a directly measurable trace of the dynamical wind-wave coupling, possible from remote sensing platforms.

\section{Analysis of wave-breaking measurements}
\label{sec:data}

\begin{figure}
   \centering
    \includegraphics[width=\textwidth]{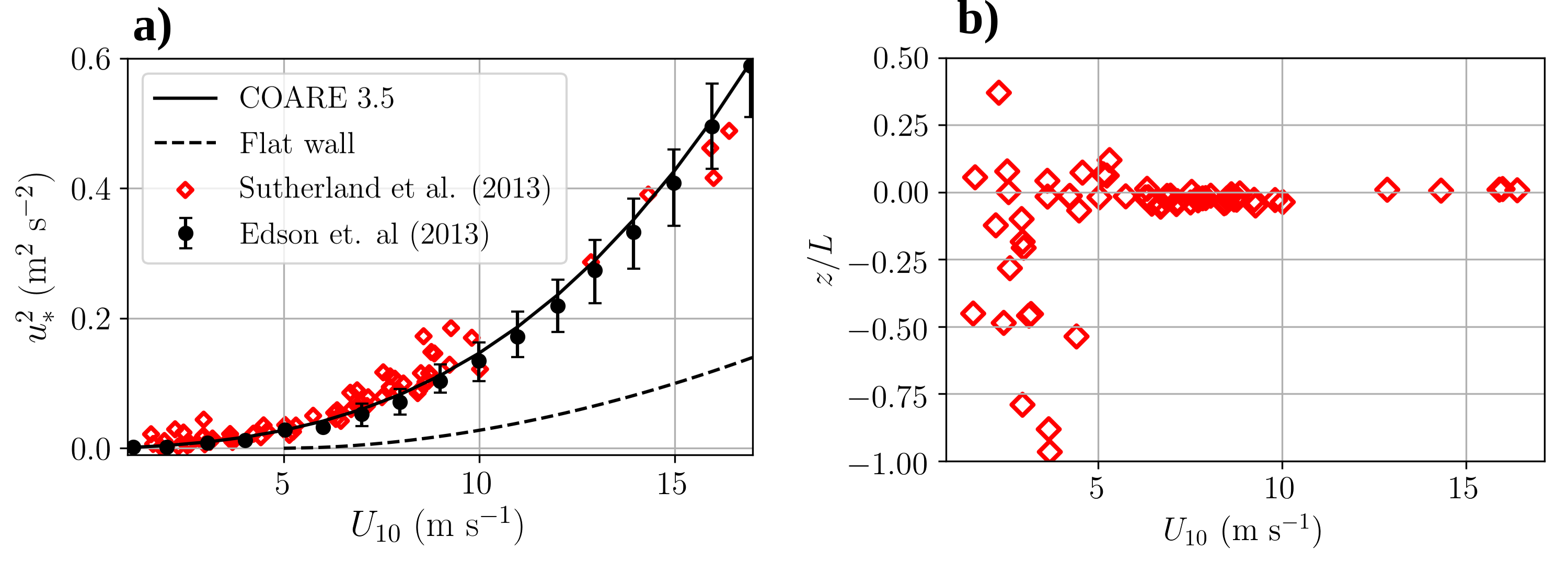}
    \caption{(a) Momentum flux $u_*^2$ versus $U_{10}$ from the COARE 3.5 parameterization (solid line) and the corresponding open-ocean measurements  \citep[dots and bins, from][]{edson2013exchange}. The relation derived for a smooth-wall boundary layer is included for reference (dashed line). The open-ocean measurements used in this study \citep{sutherland2013field, sutherland2015field} are shown as diamonds. In \S\ref{sec:pheno}, only the solid line is used in the turbulence phenomenological model to obtain the representative breaker speed. (b) Atmospheric stability versus 10-m wind speed for the open-ocean measurements used in this study. Atmospheric stability was quantified by the dimensionless stability parameter $z/L$, where $z$ is the height of the measurement, and $L$ the Obukhov length. Only measurements for which  $-0.2 \leq z/L \leq 0.2$ have been used in \S\ref{sec:data}.}
    \label{fig:measurements}
\end{figure}

\subsection{Data}
Open-ocean measurements of wave-breaking statistics from \citet{sutherland2013field, sutherland2015field} were used. The measurements were obtained during three field campaigns on-board R/V \emph{FLIP}: the RaDyO experiment, which took place 120 km south of the Island of Hawai’i in 2009 \citep{dickey2012introduction}, and the HiRes and SoCal experiments, which took place 25km off the coast of Northern California and in the Southern California Bight, respectively. As shown in figure~\ref{fig:measurements}a (red diamonds), the 10-meter mean wind $U_{10}$ (averaged over 20-minute intervals) ranged from 2 to 16 m s$^{-1}$ with few measurements for $U_{10} > 10$ m s$^{-1}$. Figure~\ref{fig:measurements}a also shows the measured friction velocity $u_*$  defined by $u_*^2 = \tau/\rho_a$, where $\tau$ is the measured turbulent stress at $10$-m height and $\rho_a$ is the mean air density.  These measurements are consistent with other experiments \citep[summarized in][dots and bins]{edson2013exchange} used to calibrate the widely used COARE parameterization (solid line). 
The presence of wind-waves at equilibrium with the wind results in a bulk drag coefficient $C_d=u_*^2/U_{10}^2$ that is higher than a $C_d$ associated with a flat-solid surface \citep[dashed line, and see also e.g.][their figure~3]{ayet2020impact}.
Except for very low winds, atmospheric stability was near neutral during the measurements (see figure \ref{fig:measurements}b). For this reason, atmospheric stability effects are not discussed further and low $U_{10}$ runs are filtered out so as to only maintain near-neutral atmospheric stability conditions. In the measurements, no strong correlation was found between variations in atmospheric stability conditions for low winds and variations in $u_*$ for a given $U_{10}$.

To each point in figure~\ref{fig:measurements}a corresponds a $20$-minute averaged spectrum of wave-breaking statistics, measured using a stereo pair of long-wave infrared cameras. More precisely, the stereo images permit the estimation of the length and speed of the breaker fronts \citep{sutherland2013field}. These measurements can be used to compute an effective $20$-minute averaged spectrum $\Lambda(c)$ of the breaker front length per unit area of sea surface per unit increment of breaking front velocity $c$, which we analyse in the following. In contrast to measurements in the visible range \citep[][]{kleiss2010}, the use of infrared cameras allows the measurement of non-air-entraining micro-breakers with crest speeds as slow as the gravity-capillary phase speed minimum (corresponding to a wavelength of 1.7 $\times 10^{-2}$ m).

\subsection{Analysis and results}

\begin{figure*}
   \centering
    \includegraphics[width=\textwidth]{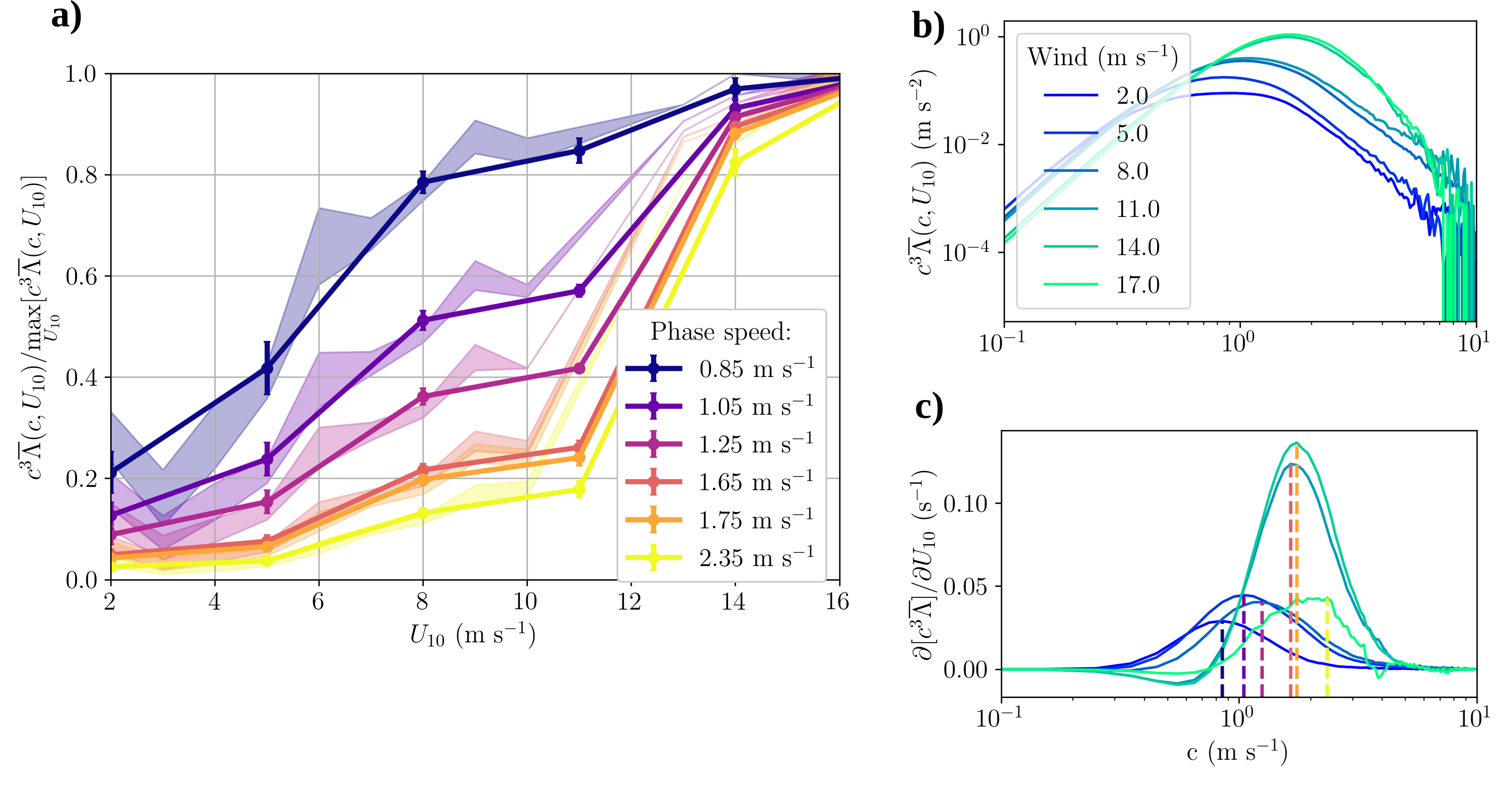}
    \caption{Analysis of the wave breaking measurements. (a) Average and standard deviation of $c^3 \Lambda(c)$ for $U_{10}$-bins, for different breaker speeds, vs $U_{10}$  (the $U_{10}$-values correspond to the middle of the bin). Dots and error bars correspond to data binned in $U_{10}$ intervals of 3 m s$^{-1}$, and shadings to intervals of 1 m s$^{-1}$. For $U_{10}$-bins  of $3$ m s$^{-1}$, (b) binned  $c^3\overline{\Lambda}(c, U_{10})$ vs breaker speed for different $U_{10}$. and (c) gradient of the binned $c^3 \overline{\Lambda}(c, U_{10})$ with respect to changes in $U_{10}$ vs breaker speed, for different $U_{10}$. The vertical dashed lines correspond, for each bin, to the breaker speed whose gradient $\partial[c^3 \overline{\Lambda}(c, U_{10})]/\partial U_{10}$ is the largest, with colours consistent with (a).}
    \label{fig:lambdas}
\end{figure*}

The statistics of the breaker front length $\Lambda(c)$ were originally introduced as a measurable quantity to characterise the wind-and-waves equilibrium \citep{phillips1985spectral}. Here, we argue that the third moment of this distribution ($c^3\Lambda(c)$) is the appropriate quantity to characterise air-sea momentum fluxes. Originally related to mass fluxes at the air-sea interface \citep{deike2017air, deike2018gas}, $c^3 \Lambda(c)$ can also be linked to the roughness elements generated by breaking waves: near breaking crests, local energy fluxes disrupt the sea surface and generate a spectrum of roughness elements (i.e. roughness elements at different horizontal scales).  \citet{newell1992rough} suggested that these roughness elements are at a Kolmogorov-type equilibrium.  This equilibrium requires that the production rate, the energy transfer rate across scales, and the dissipation rate be balanced and scale-independent (i.e. a conservative cascade). Based on this idea,  \citet{Kudryavtsev2014} showed that this energy flux across scales should be proportional to $c^3$ (where $c$ is the phase speed of the breaker front generating the roughness elements). The authors finally demonstrated that this spectrum of roughness elements modifies near-surface turbulent stress and that the intensity of the modification depends on $c^3\Lambda(c)$. Hence $c^3 \Lambda(c)$ is physically related to the effect of breaking waves on near-surface turbulence. 

To analyse this effect further, data were first binned in $U_{10}$-intervals of 3 m s$^{-1}$. Dots and error bars in figure~\ref{fig:lambdas}a show the resulting $U_{10}$-dependent average $c^3\overline{\Lambda}(c, U_{10})$ and standard deviation for each bin and different breaker speeds, normalised by the $c$-dependent maximum  of $c^3\overline{\Lambda}(c, U_{10})$ with respect to $U_{10}$. Examples of the dependence of $c^3\overline{\Lambda}(c,U_{10})$ with $c$ are also shown in figure~\ref{fig:lambdas}b for reference. Figure~\ref{fig:lambdas}a shows that, for a given breaker speed, the increase in $c^3\overline{\Lambda}(c, U_{10})$ is not constant with increasing wind speed, but seems be strong around a particular wind speed before reaching a saturation value: this is mostly apparent for the blue line, corresponding to a breaker speed of $0.85$ m s$^{-1}$. To further quantify this effect, i.e.  how \emph{changes} in $U_{10}$ leave imprints on the local wind-waves breaking field, what is analysed in the following is the \emph{gradient} of $c^3\overline{\Lambda}(c, U_{10})$ with respect to $U_{10}$ as shown in figure~\ref{fig:lambdas}c for different wind speeds.

Figure~\ref{fig:lambdas}c reveals that, for a given $U_{10}$ this gradient ($\partial [c^3\overline{\Lambda}(c, U_{10})]/\partial U_{10}$) is found strongest around a speed $c_r$, termed representative breaker speed in the following (vertical lines in figure~\ref{fig:lambdas}b). This speed increases with $U_{10}$: for $0.5 < U_{10} < 3.5 $ (m s$^{-1}$), it is $c_r = 0.85$ m s$^{-1}$ (blue vertical line in figure~\ref{fig:lambdas}c) while for $12.5 < U_{10} < 15.5 $ (m s$^{-1}$), it is $c_r =1.75$ m s$^{-1}$ (orange vertical line in figure~\ref{fig:lambdas}c). Note that, even though for  $15.5 < U_{10} < 18.5 $ (m s$^{-1}$) such a wave speed can still be defined  ($c_r = 2.35$ m s$^{-1}$, yellow line in figure~\ref{fig:lambdas}c), the low number of measurement points contained in the bin induces noise in the gradient of $c^3 \overline{\Lambda}(c, U_ {10})$. 

Hence, for each $U_{10}$ interval, a representative breaker speed $c_r$ can be defined, corresponding to the short-scale waves whose breaking statistics  ($c^3\Lambda(c)$) are the most sensitive to changes in mean wind speed (solid blue line in figure~\ref{fig:scalewise}). Sensitivity to the bin size was tested using 10-m wind speed binning intervals of a width of 1 and 2 m s$^{-1}$. The resulting variability of the average value of $c^3\Lambda(c)$ in each bin is shown as shadings in figure~\ref{fig:lambdas}c, and as dashed and dotted lines in figure~\ref{fig:scalewise}. The outcomes are qualitatively consistent with the analysis above. 

Finally, a sensitivity analysis on the choice of the moment $n$ of the wave breaking distribution $c^n\overline{\Lambda}(c, U_{10})$ was also performed. Indeed, even though there is a physical reason to the choice $n=3$, other exponents could have been used to extract a representative breaker speed from the data. As explained in appendix \ref{app:sensitivity}, the exponent $n=3$ used above is the one for which the representative speed $c_r$ is the most unambiguously defined. This choice is also justified by the agreement with the turbulence model (see \S\ref{sec:pheno}), and by the tri-dimensional view of eddies filling the space between waves presented in \S\ref{sec:tri}.

\subsection{A measurable kinematic relation}\label{subsec:kinematic}
A key property of the representative breaker speed $c_r$ is its kinematic link with the friction velocity 
\begin{equation}\label{eq:speed-friction-wave-relation}
    c_r = 2.5 u_*.
\end{equation}
While an approximate form of this relation can be obtained from a best fit to the data (in figure~\ref{fig:scalewise}), the turbulence model presented in the next section provides an exact derivation of (\ref{eq:speed-friction-wave-relation}). 

Such a relation was already discussed in \citet[p.~141]{phillips1966dynamics} when considering the interaction between small scale waves and surface drift induced by viscous wind stress: \citet{phillips1966dynamics} argued that only waves with phase speeds smaller than five times the surface drift would be strongly impacted by it. The magnitude of the surface drift was further reported to be $q = 0.55 u_*$ m s$^{-1}$ \citep{wu1975wind}. Hence, wave scales such that $c = 5 \times 0.55 u_* \sim 2.5 u_*$ are the smallest waves that do not significantly interact with surface drift. Interactions with surface drift tend to reduce the critical steepness needed for wave breaking. This leads to (i) a reduction of the amplitude of drift-affected waves with respect to a no-drift situation, and hence to (ii) a reduction of the strength of the breaking event, i.e. its capacity to be visually detected.

Consequence (ii) can explain the existence of a maximum in $\partial[c^3\overline{\Lambda}(c, U_{10})]/\partial U_{10}$ (evident in figure~\ref{fig:lambdas}c). With increasing winds, short wave-breaking occurrence is expected to increase. 
However, this increase is truncated for drift-affected waves. This leads to the Bell-shaped curves of figure~\ref{fig:lambdas}c, i.e. to the fact that a representative surface wave speed can be visually defined for a given wind: the Beaufort scale. As for consequence (i), it implies that amplitudes of drift-affected waves are reduced, and so is their contribution to the roughness elements affecting atmospheric turbulence, such as discussed in the next section.

To summarize, breaker data analysis suggests that it is possible to relate a change in $U_{10}$ to a change in the properties of breaking crests with a particular wave front velocity. This reflects both local wind stress variations and changes in wind-induced surface drift.

\section{A roughness sublayer from a phenomenological turbulence model}\label{sec:pheno}

In \S\ref{sec:data}, wave measurements were used to show the existence of a representative breaker speed linked to the local properties of surface wind. In this section, we follow the opposite reasoning by starting from atmospheric theory to find the representative breaker speed. We ask the following question: assuming that there exists a layer close to the surface where turbulence is entirely constrained by the intermittent roughness elements associated with breaking waves, what should the defining parameters of such roughness elements be? This question is addressed using a a phenomenological model of turbulence. Its original formulation, for flow above fixed roughness elements, is presented in \S\ref{subsec:pheno-model}. In \S\ref{subsec:pheno-results}, the model is extended to a windy sea, providing an independent justification for the results of the previous section, and in particular to (\ref{eq:speed-friction-wave-relation}). 

\subsection{A turbulence model above fixed roughness elements}\label{subsec:pheno-model}

The phenomenological model, originally presented in \citet{katul2011mean} following ideas of \citet{gioia2010spectral}, describes turbulence in the Surface Boundary Layer (SBL), where the flow is horizontally homogeneous and stationary with no subsidence, and hence $u_*^2$ is height-independent. Turbulence is modeled through a turbulence kinetic energy balance and so-called energy-containing eddies (red circles in figures~\ref{fig:stress}b,c), which are leading order contributors to the vertical momentum flux at a given height (dashed lines in figures~\ref{fig:stress}b,c). 

Those eddies are defined by their streamwise extension or height $s_e$ and their turnover velocity, proportional to $u_*$ \citep[e.g.][]{hunt1988turbulent}. For neutral conditions, the mean wind shear can then be expressed as
\begin{equation}\label{eq:wind-shear}
\frac{\partial U(z)}{\partial z}= \frac{u_*}{\kappa s_e(z)},
\end{equation}
where $U$ is the mean wind speed in the streamwise ($x$) direction, and $z$ is height, whose origin is the mean sea surface height, and $\kappa \sim 0.4$ is the von K\'arm\'an constant. The derivation of (\ref{eq:wind-shear}) from the model hypotheses can be found in appendix \ref{app:pheno}.

\begin{figure}
    \centering
    \includegraphics[width=0.8\textwidth]{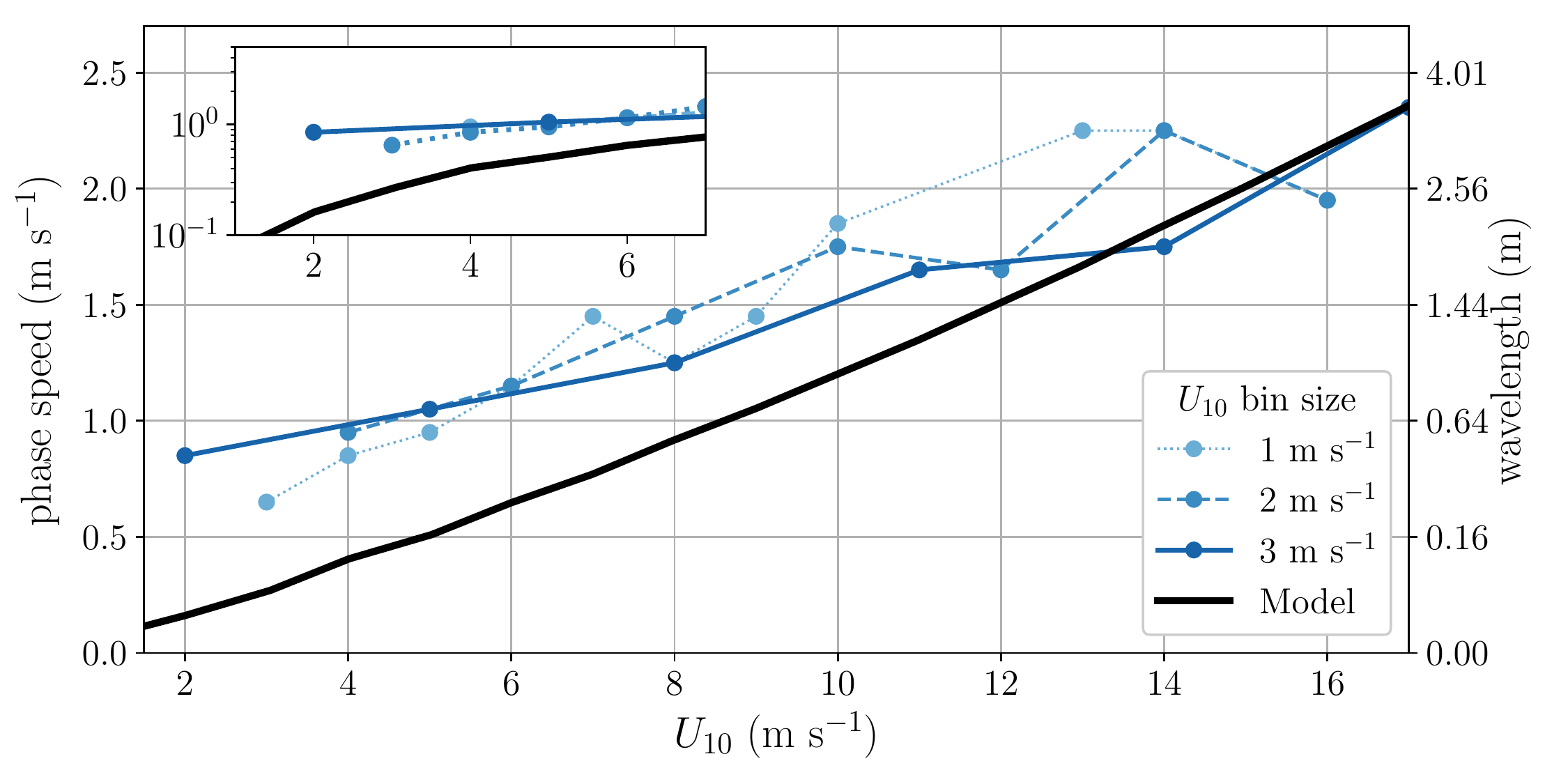}
    \caption{Representative breaker speed of the wind-wave local coupling as a function of $10$-m wind speed. Blue lines are the results from analysis of wave-breaking statistics $c^3\overline{\Lambda}(c, U_{10})$, binned in different wind intervals. Solid black line is the result from the phenomenological model of turbulence. The wavelengths ($\lambda = 2\pi c^2/g$) correspond to those of deep-water linear waves travelling with phase speed $c$, and have been included for reference only.}
    \label{fig:scalewise}
\end{figure}

The phenomenological model defined in (\ref{eq:wind-shear}) thus links the mean wind shear to the friction velocity $u_*$ and the size of the energy-containing eddies $s_e$. For flow past a rough wall, \citet{gioia2001scaling} and \citet{bonetti2017manning} proposed that the surface boundary can be decomposed into two sublayers, depicted in figure~\ref{fig:stress}a. The highest sublayer is a logarithmic layer, in which the size of the energy-containing eddies scales with distance from the wall ($s_e(z)= z$), corresponding to attached eddies \citep[as proposed in][and depicted in figure~\ref{fig:stress}b]{townsend1980structure}. From (\ref{eq:wind-shear}), the mean wind profile is hence logarithmic
\begin{equation}\label{eq:log-law}
U_{\mathrm{log}}(z) = \frac{u_*}{\kappa} \log(z/z_0), \;\; \mathrm{for }\; z \geq h_r,
\end{equation}
where $z_0$ is the roughness height. Below the logarithmic sublayer lies the roughness sublayer of height $h_r$. In this layer, the size of the energy-containing eddies scales with the height of the roughness elements that is proportional to $h_r$. To ensure continuity of $\partial U/\partial z$ at the interface between both layers, the scaling constant is considered to be equal to one, i.e. $s_e(z) = h_r$ for $z \leq h_r$  \citep[see figure~\ref{fig:stress}c and][]{ bonetti2017manning}. Equation (\ref{eq:wind-shear}) then yields a linear wind profile 
\begin{equation}\label{eq:linear-law}
U_{\mathrm{lin}}(z) = \frac{u_*}{\kappa} \frac{z}{h_r}, \; \; \mathrm{for } \;  0 \leq z \leq h_r .
\end{equation}
The height of the roughness elements $h_r$ is different from the roughness height $z_0$ (which is derived from an extrapolation of the logarithmic wind profile to $U=0$) and, by definition, $h_r \geq  z_0$ \citep{raupach1999rough}. 

Finally, a relation between $z_0$ and $h_r$ can be obtained by requiring continuity of $U$ at $z=h_r$, and reads
\begin{equation}\label{eq:z0-hr}
z_0 =  h_r\exp(-1).
\end{equation}
In the aerodynamically smooth regime, $h_r$ is the height of the viscous sublayer, $\exp(1) z_0^v$, where $z_0^v =  \gamma \nu/u_*$ is the viscous roughness height (and $\nu$ and $\gamma \sim 0.11$ are the kinematic viscosity of the air and the roughness Reynolds number of smooth flows, respectively). For this regime, the resulting relation between $u_*^2$ and $U_{10}$ is plotted in figure~\ref{fig:measurements}a (dashed line).

\begin{figure}
    \centering
\includegraphics[width=0.6\textwidth]{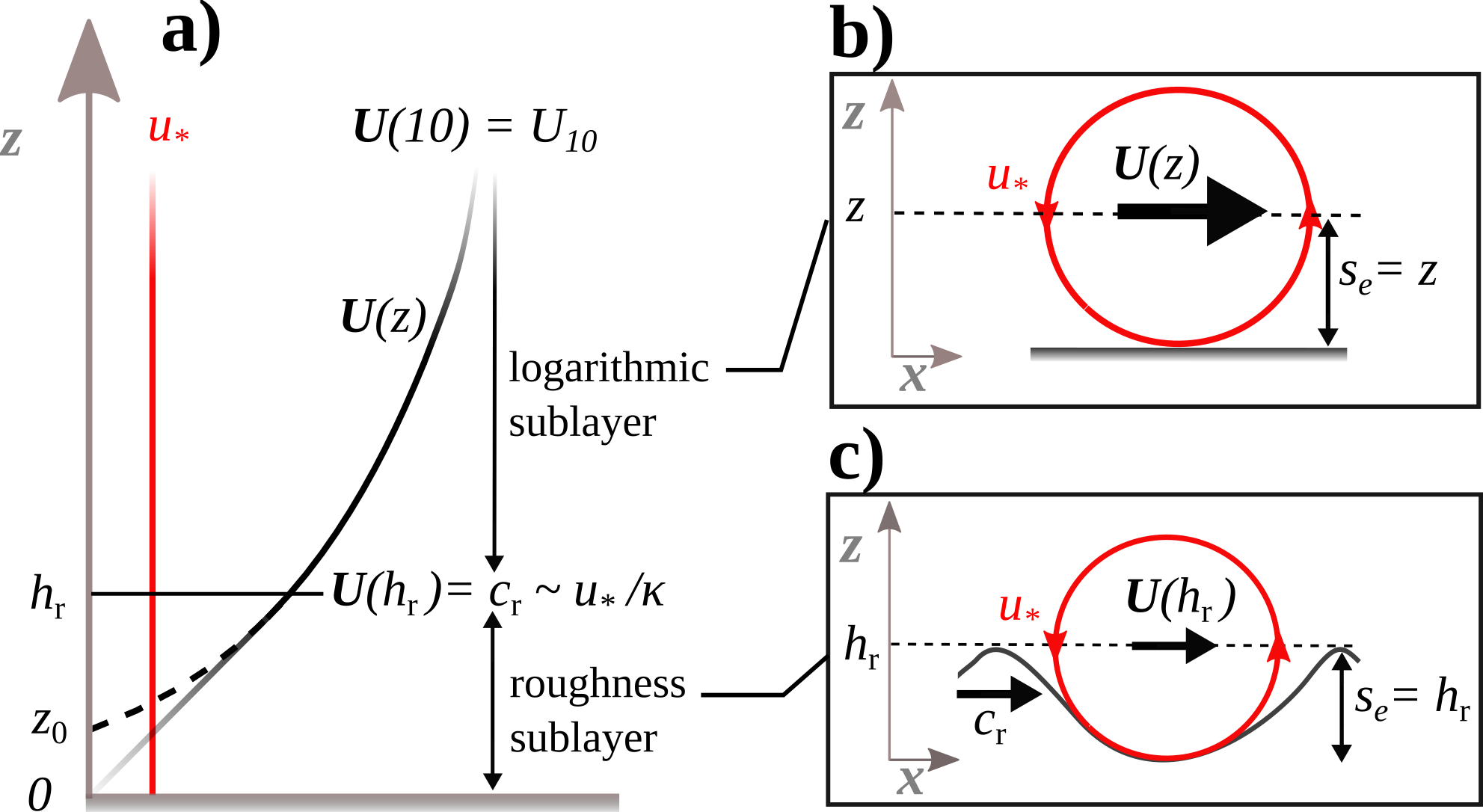}
\caption{Phenomenological model describing the surface boundary layer (SBL) in the presence of breaking waves. (a) The SBL is divided into two sublayers: (b)  the logarithmic sublayer, where energy-containing eddies scale with $z$ (attached eddies); (c) the roughness sublayer, where their size is independent of height and set to $h_r$, the \emph{effective height} modeling the intermittent effect of breaking waves. In the roughness sublayer above waves, no mean wind profile is prescribed, contrary to the case above fixed roughness elements where it is linear \citep{bonetti2017manning}.}
    \label{fig:stress}
\end{figure}

\subsection{Model extension to an ocean surface with breaking waves}\label{subsec:pheno-results}

An ocean surface with breaking waves of multiple scales and velocities is now considered. We assume that the SBL can still be decomposed into two sublayers. In the logarithmic sublayer, the conventional picture of attached eddies holds. In the roughness sublayer, we seek the measurable properties of the roughness elements that set the geometry of the energy-containing eddies. To do so, we adopt a different interpretation of the roughness sublayer than above fixed roughness elements \citep{bonetti2017manning}.

First, we do not assume a linear wind profile in the roughness sublayer. Instead, we only require the bulk mean wind shear across the roughness sublayer to follow
\begin{equation}\label{eq:bulk-wind-shear}
\frac{U(h_r)}{h_r} = \frac{u_*}{\kappa} \frac{1}{h_r} .
\end{equation}
This is a weaker assumption than requiring the validity of (\ref{eq:wind-shear}) at all heights (and hence the linear wind profile (\ref{eq:linear-law})). This assumption hence covers the occurrence of more complex wind profiles that can occur close to breaking waves, within the roughness sublayer. 
%
In fact, the mean wind speed in the roughness sublayer does not have a well-defined physical interpretation. In the logarithmic sublayer, $U(z)$ can be interpreted as the convection velocity of energy-containing eddies of size $s_e = z$ \citep[see e.g.][]{phillips1957generation} while, in the roughness sublayer, the convection velocity could be ill-defined due to transient roughness elements \citep{kraus1967wind}.
Certainly, it is more reasonable to assume that the convection velocity of eddies at all heights below $h_r$ is $U(h_r)$, i.e. the convection velocity of the roughness-sublayer eddies of size $s_e = h_r$.

The second difference with respect to flow over fixed roughness elements is the interpretation of $h_r$. We interpret $h_r$ as being an \emph{effective} height of the roughness elements, modeling the bulk effect of intermittent wind-wave breaking events on the SBL. It does not correspond to a measurable height (e.g. a monochromatic wave), but is instead only measurable through its dynamical link with $c_r$. 

This link results from the dynamical relation between the bulk wind shear and the properties of energy-containing eddies in the roughness sublayer. Energy-containing eddies can indeed be seen as resulting from the interaction between mean wind shear and turbulence: rearranging (\ref{eq:wind-shear}), the streamwise extension of energy-containing eddies (which defines their properties) can be expressed as $s_e \propto w_eT_e$, i.e. as a product of an eddy turnover velocity $w_e \propto u_*$ and time $T_e \propto  (dU/dz)^{-1}$.
For roughness-sublayer eddies, the turnover velocity further reads $T_e = (U(h_r)/h_r)^{-1}$, indicating that those are controlled by the \emph{bulk} mean wind shear over the sublayer (defined in (\ref{eq:bulk-wind-shear})). Roughness-sublayer eddies can thus be viewed as originating from an instability of a similar form than above forest canopies \citep{raupach1996coherent}, for which eddies have a size $s_e$ proportional to the scale of the wind shear $h_r$ \citep[this scale is also called the ``vorticity thickness" in e.g.][]{harman2007simple}. Now, if the wind shear in the roughness sublayer is set by intermittent yet intense wave breaking events, fluid is entrained at the speed $c_r$ of the breaker fronts on top of the roughness sublayer \citep[as suggested by][]{melville1977wind}. Hence $U(h_r)$, the magnitude  of the bulk wind shear across the roughness sublayer should be equal to the wave speed, i.e.
\begin{equation}\label{eq:match-wind-wave}
U(h_r) = c_r = u_*/\kappa,
\end{equation}
providing a theoretical justification to (\ref{eq:speed-friction-wave-relation}).  

It should be stressed that the argument leading to (\ref{eq:match-wind-wave}) relies on two conditions. First, it requires that at the top of the roughness sublayer $h_r$, $U(h_r)$ is non-zero and is physical. Hence $z_0$ could not have been chosen as a roughness sublayer height, since it is the extrapolated height where the wind speed $U(z_0)$ is zero for a logarithmic wind profile. Second, a logarithmic profile for $z > h_r$ is required to obtain (\ref{eq:match-wind-wave}). This second point can be questioned since, for $z> h_r$, the presence of waves and wave growth induces wave-induced motions, which cause a deviation of the mean wind from a logarithmic profile \citep[e.g.][]{miles1965note}. In appendix \ref{app:extend}, the phenomenological model is extended to account for these motions above the roughness sublayer \citep[following][]{ayet2020impact}. In this case, equation~(\ref{eq:z0-hr}) is more complex, but the resulting $c_r$ is of the same order of magnitude than (\ref{eq:match-wind-wave}) (see figure \ref{fig:s:height}c). 
    
The extension of the phenomenological model also revealed that as wind increases and wave-induced motions become more energetic, the ratio $s_e/h_r$ increases (up to 2.2 for 16 m s$^{-1}$, while it is one in the absence of wave-induced motions, figure \ref{fig:s:height}b). This finding is consistent with detailed laboratory experiments of flow over granular roughness elements \citep{manes2007double} that reported a log region commencing at approximately 1.6 times the mean grain height, coincident with the turbulent flow statistics attaining spatial uniformity (i.e. the effect of the roughness heterogeneity is no longer felt). The ratio $s_e/h_r$ being larger than one also supports the interpretation of the roughness elements associated to breaking waves as being spatially and temporally intermittent, with an average size ($h_r$) smaller than the scale ($s_e$) of the eddies they generate. Overall, these findings reveal the existence of a direct interaction between wave growth and roughness-sublayer eddies, through the presence of wave-induced motions (which are perturbations of the airflow, coherent with wind-waves, at heights $z > h_r$). 

To summarise, the phenomenological model used here does not contain \emph{any} a priori spectral information on the wave field, and solely requires specification of the bulk information on turbulence contained in open-ocean measurements of $u_*$ and $U_{10}$ (e.g. the measurements presented in figure \ref{fig:measurements}a, solid line). Yet, the resulting wave velocity is similar to the one obtained from analysis of independent wave-breaking statistics. This is shown in figure~\ref{fig:scalewise}, where the wave velocity $c_r$ corresponding to the velocity obtained from (\ref{eq:match-wind-wave}) (black line) is of the same order of magnitude than that of obtained from data (blue lines), with a similar trend as a function of $U_{10}$. For very low winds, model and data trends disagree (see inset of figure~\ref{fig:scalewise}), but the resulting breaker phase speed is close to the capillary-gravity transition, and hence also to the measurement limit of the wave-breaking data. 

This analysis shows that there is a correspondence between the properties of the wave breaking field and those of the energy-containing eddies. This positive result is used below to propose a three-dimensional interpretation of the phenomenological model in the presence of roughness elements associated to wave breaking.

\section{A three-dimensional view of wave-breaking-constrained turbulence}\label{sec:tri}

\begin{figure*}
    \centering
\includegraphics[width=0.8\textwidth]{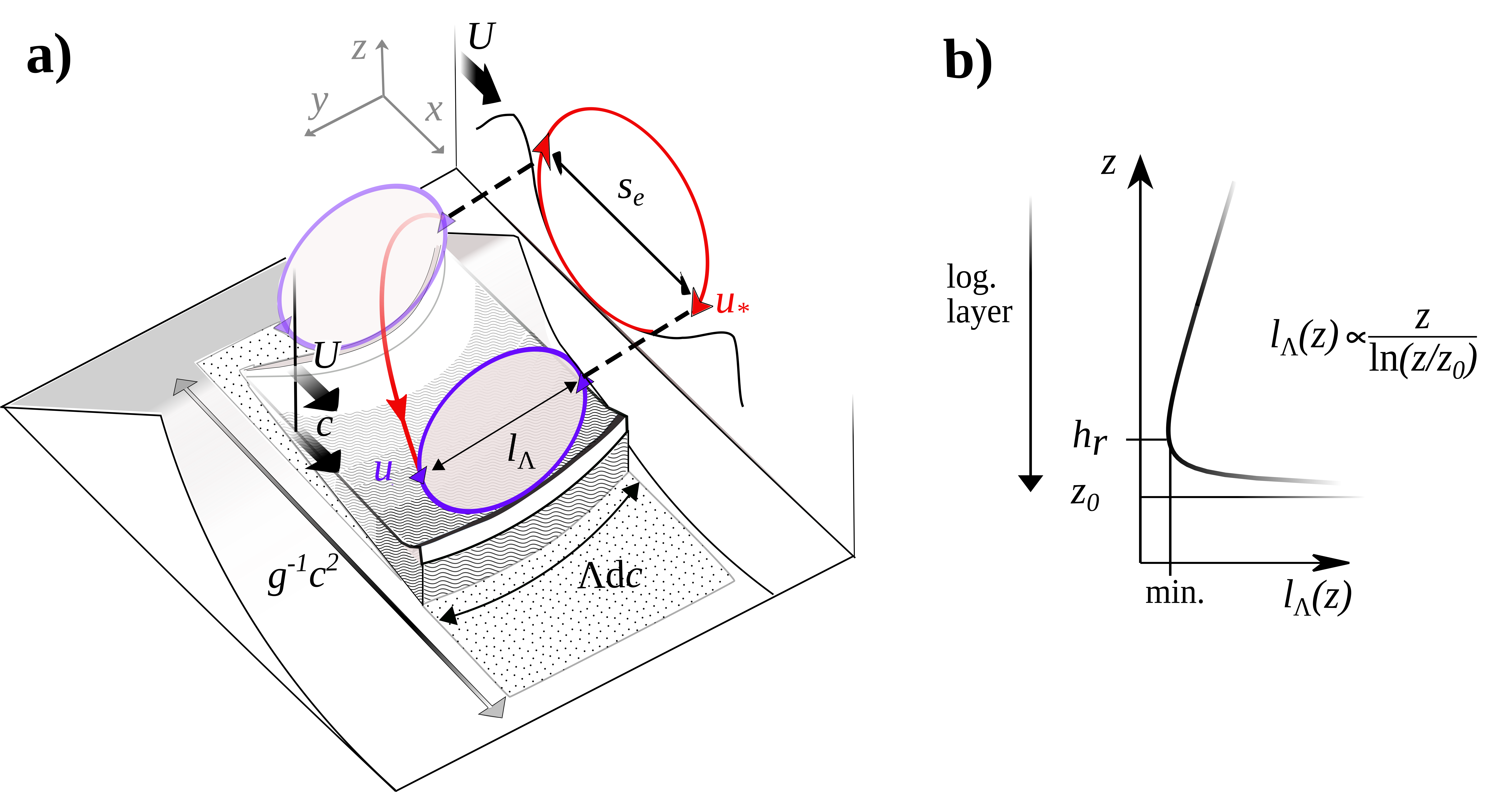}
\caption{Three-dimensional interpretation of the phenomenological model (presented on figure~\ref{fig:stress}) over a rough sea surface. (a) Breaker fronts of speed $c$ and spanwise extension $\Lambda dc$ constrain the spanwise extension $l_\Lambda$ of vortices advected at a speed $U = c$ (purple circles). The streamwise scale of energy-containing eddies $s_e$ (red circle) results from the spiraling motion of a fluid parcel (red line) due to advection at the mean wind speed and rotation at a turnover velocity $u_*$. Over their liftetime (longer than the eddie lifetime), breaker fronts cover an approximate streamwise distance $g^{-1} c^2$, the dotted area. Note the difference in the geometry of the individual breaker front and of the effective roughness element (black line in the $(x,z)$ transect). (b) Required $l_\Lambda$ for logarithmic-layer energy-containing eddies, i.e. for $s_e = z$ and $U \propto u_* \ln(z/z_0)$.}
    \label{fig:3D}
\end{figure*}{}

In the previous section, a roughness height was determined and was related to the streamwise scale of energy-containing eddies $s_e$. Theoretical \citep{csanady1985air, eifler1993hypothesis}, experimental \citep{reul1999air} and numerical \citep{suzuki2011turbulent} work further revealed that the interaction between breaking waves and turbulence is a three-dimensional processes in which the spanwise length of the breaking crests plays an essential role. In the following, the phenomenological model of turbulence is extended so as to relate this spanwise length to the streamwise scale of energy-containing eddies. The model is sketched in figure~\ref{fig:3D}a and is described below. The spanwise direction is labelled as $y$.

Defining $l_\Lambda$ to be a spanwise extension of surface roughness elements (figure~\ref{fig:3D}a), $l_\Lambda$ is then proportional to the length of the individual crests associated with breaking fronts of speed $c$, and the sum of the latter, per unit area of ocean surface, is $\Lambda(c)dc$.  Roughness elements are assumed to trigger spanwise atmospheric eddies of similar extension with a turnover velocity proportional to $u_*$.  These eddies are advected at the mean wind speed $U$ (compare the two purple circles in figure~\ref{fig:3D}a). The turnover time of the eddies is $T_\Lambda \propto l_\Lambda/u_*$, and the streamwise distance, $s_e$, traveled by a representative fluid parcel during a turnover time $s_e = U T_\lambda$, is
\begin{equation}\label{eq:link_spanwise}
s_e(z) \propto U(z) l_\Lambda(z)/u_*. 
\end{equation}

As sketched in figure~\ref{fig:3D}a, the distance $s_e$ can be represented as the streamwise Lagrangian distance separating the ascending and descending branches of the eddy, in between which a representative fluid parcel would undergo a spiraling motion. It is also the streamwise extension of the energy-containing eddies defined in the phenomenological turbulence model (red circle in figure \ref{fig:3D}a). As for the phenomenological model, this representation is to be understood in an ensemble mean sense, unlike similar but instantaneous representations of turbulence over flat walls \citep{kline1967structure, blackwelder1979streamwise} and waves \citep{csanady1985air, eifler1993hypothesis} that would result from conditional sampling of the flow.

This three-dimensional interpretation relates, through (\ref{eq:link_spanwise}), the streamwise extension of the energy-containing eddy $s_e$ to the spanwise extension of the roughness elements constraining its size $l_\Lambda$. For a logarithmic layer, for which $s_e = z$ and $U \propto u_* \ln(z/z_0)$, equation (\ref{eq:link_spanwise}) then yields the height-dependent $l_\Lambda(z)$ required to obtain attached energy-containing eddies at each height $z$ satisfying the law of the wall. As shown in figure~\ref{fig:3D}b, after some elementary algebra, $l_\Lambda(z)$ has a minimum, reached at a height corresponding to the roughness sublayer height $h_r$ defined in (\ref{eq:z0-hr}). For a wavy surface, an increase of $l_\Lambda$ with decreasing height is not physical (as happens for $z < h_r$). This would indeed imply an increase of the spanwise extension of breaking fronts (proportional to $l_\Lambda$) as height decreases. Noting that, for decreasing heights, the speed (and  wavelength) of the representative breakers decreases (from~(\ref{eq:match-wind-wave})), this would imply an increase in the breaking-front average length with decreasing wavelength, which is not realistic \citep[see][which, among others, highlight the self-similarity of breaking fronts]{belcher1997breaking, reul2003model}. Hence, breaking fronts can only imprint their spanwise scale on energy-containing eddies for $z \geq h_r$.  

Subsequently, $h_r$ represents the smallest height at which roughness elements associated with breaking fronts can set the spanwise and streamwise extension of energy-containing eddies. Below this height, the scale of energy-containing eddies is  $h_r$. Furthermore, $h_r$ is proportional to the spanwise size of the roughness elements associated to breaking fronts ($l_\Lambda$). This interpretation offers an additional argument for the importance of $h_r$ in the characterization of the wind-over-waves coupling.

From the wave-breaking data, the scale $h_r$ emerged from the variations of $c^3 \overline{\Lambda}(c, U_{10})dc$ with $U_{10}$. Assuming foam that patches have a lifetime proportional to the underlying wave period \citep{reul2003model}, $g^{-1} c^2\Lambda(c)dc$ is related to the the fraction of sea-surface turned over by breaking fronts, weighted by their lifetime. Hence $c^3\Lambda(c)dc$ also contains information about both the lifetime of roughness elements and their momentum (proportional to $c$). The fact that a change in the properties of roughness-sublayer energy containing eddies is coupled to a change in $c_r^3\Lambda(c_r)dc$ thus indicates that not only the size of the breaking fronts ($\propto l_\Lambda$) but also their momentum ($\propto c_r$) and lifetime ($\propto c_r/g$) are needed for the description of near-surface turbulence properties. 

\section{Concluding discussion}\label{sec:discussion}

\begin{figure}
    \centering
    \includegraphics[scale=0.4]{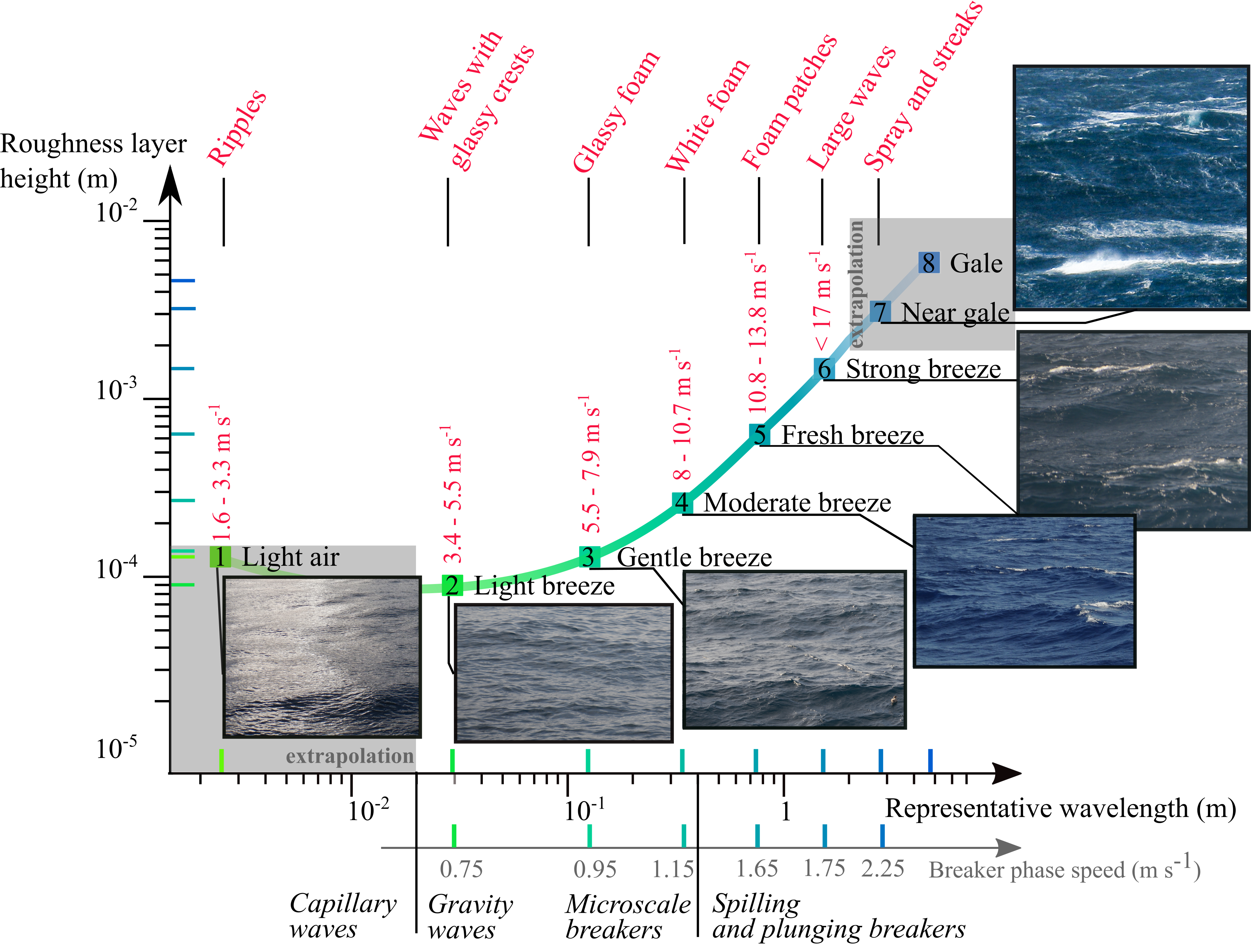}
    \caption{Relation between the two wind-dependent quantities proposed in this work and the Beaufort scale (red text and curve, with speeds corresponding to $U_{10}$ intervals): a representative speed of breaking fronts (horizontal axis, converted to a wavelength for reference), and the height of the roughness sublayer, equivalent to the smallest size of energy-containing eddies (the vertical axis). Note that the classification of breaker types on the horizontal axis \citep[from][]{katsaros1992dependence} corresponds to the qualitative sea-surface description related to the Beaufort scale (red text).}
    \label{fig:beaufort-summary}
\end{figure}

In this work, we identified a surface speed $c_r$ corresponding to small scale breakers that are most coupled to the atmosphere. Infrared measurements of wave breaking statistics have indeed revealed that variations of wave breaking statistics ($c^3 \overline{\Lambda}(c, U_{10})$)  with ten-meter wind are significant only for a particular breaker speed $c_r$, which increases with friction velocity (equation (\ref{eq:speed-friction-wave-relation})). Following the ideas of the Beaufort scale, breakers travelling at this speed can be interpreted as being the most visually related to a particular near-surface wind speed. This is shown in figure~\ref{fig:beaufort-summary}, where the visual characteristics of the surface associated to the Beaufort scale (red text) are compared to the breaker speed identified in this analysis and its classification following \citet[][italic text in the figure]{katsaros1992dependence}.

We then extended a phenomenological turbulence model to interpret this measurable relation as a macroscopic property of the near-surface turbulent air flow. Without specifying any information on the wave spectrum, we found that the breaker speed $c_r$ is associated with the sea-surface roughness elements that affect the most energy production of near-surface turbulence eddies. This creates a roughness sublayer in which the scale of near-surface eddies is constrained by the breaker speed $c_r$ (\S\ref{sec:pheno}), but also (as shown in \S\ref{sec:tri}) by its spanwise length and lifetime (i.e. the dotted area in figure \ref{fig:3D}). The height $h_r$ of the roughness soublayer is small, ranging from $10^{-4}$ m to $10^{-2}$m (see figure \ref{fig:beaufort-summary}). 


At first glance, the roughness sublayer height could be interpreted as (i) the height at which airflow separation events induced by waves of phase speed $c_r$ affect near-surface turbulence \citep{kraus1967wind, melville1977wind}, or (ii) the height at which inviscid instabilities above a free surface are generated \citep{miles1957generation}. While both mechanisms certainly play a role in the generation of roughness-sublayer eddies, they are associated with wind-wave generation, which is not what we aim to describe through roughness-sublayer energy-containing eddies. The latter should instead be viewed as wave-coherent motions \citep[as defined in][]{stewart1961wave}, which are an indirect imprint of wind-wave growth on atmospheric turbulence. The imprint is indirect since: (i) wind-wave growth occurs due to the coupling between waves slower than $c_r$ and turbulence at heights greater than $h_r$ \citep[][see also appendix \ref{app:extend}]{Kudryavtsev2014, ayet2020impact} (ii) this coupling and wave growth then sets the statistics of breaking waves \citep{phillips1985spectral}, (iii) this finally affects roughness-sublayer eddies.


While the existence of a roughness sublayer over a windy sea has already been suggested elsewhere \citep[e.g.][]{kraus1967wind,kitaigorodskii1973physics, csanady1985air}, the properties of roughness-sublayer eddies proposed herein are not supported by empirical evidence, but based on an analogy with other types of surfaces \citep{raupach1988canopy,bonetti2017manning}. Importantly,  energy-containing eddies have been related to the spectral properties of turbulence in prior work \citep{katul2014cospectral}. This enables the new hypotheses on wind-wave interactions formulated in this work to be tested using (i) direct numerical simulations \citep{yang2009characteristics}, which could leverage the spectral link between energy-containing eddies and the turbulent pressure-strain correlations \citep{ayet2020scalewise} or (ii) in situ measurements  \citep{ortiz2019evaluation}, to directly analyze the shape of the turbulence spectrum \citep[related to energy-containing eddies, see][]{ayet2020scaling}.

Finally, the proposed breaker speed $c_r$ is a more general descriptor of the wind-and-waves dynamical coupling than the roughness height $z_0$. The speed $c_r$ can be determined from direct measurements without a priori specification of turbulence or wave properties. We anticipate that $c_r$ is sensitive to variations in parameters affecting the wind-and-waves coupling (besides variations of $U_{10}$ already captured by $z_0$). Indeed, $c_r$ results from the selective attenuation of the amplitude of short waves by surface drift (as discussed in \S\ref{subsec:kinematic}). The amplitude of those waves is generally described by a wave action budget balancing wind input, breaking wave dissipation and non-linear wave-wave interactions \citep{phillips1985spectral}. On the other hand, the size of roughness-sublayer energy-containing eddies is also dependent on the presence of wave-induced motions (appendix \ref{app:extend}). Hence, the dynamical wind-wave coupling expressed by the representative velocity should be sensitive to modifications of these different processes by environmental conditions, e.g. the presence of slicks, surface currents and modulating longer waves that can alter surface drift \citep{phillips1974wave}, the wave action budget \citep{kudryavtsev2005radar} and wave-induced motions \citep{ayet2020impact}. Additionally, atmospheric stability effects, not considered in the present work, can be integrated into the proposed phenomenological framework, as modifying the scale of energy-containing eddies \citep{katul2011mean,li2012mean}. Assessing the sensitivity of $c_r$ to these parameters will be the focus of future work. 

More generally, $c_r$ captures the effect of a multiscale moving ocean surface on atmospheric turbulence and its modification due to external parameters. It is a promising measurable candidate for the characterization of air-sea interactions both by in situ or remote sensing methods. Measuring $c_r$ is challenging and would imply capturing a characteristic velocity (associated to a breaker scale) that changes depending on wind speed (and probably on the external parameters mentioned above). As such, this can guide methods for the interpretation of remote sensing observations, sensitive to breakers and foam-coverage properties, as well as help design future satellite-borne instruments, especially to directly retrieve sea surface Doppler estimates.

\section*{Declaration of Interests}
The authors report no conflict of interest.

\section*{Acknowledgments}

The code used to generate the figures is available on Github [LINK WILL BE PROVIDED UPON ACCEPTANCE]. AA  was supported by DGA grant No D0456JE075, the French Brittany Regional Council and ANR grant ANR-10-IEED-0006-26   .  PS was supported by funding from the European Research Council (ERC) under the European Union’s Horizon 2020 research and innovation programme (grant agreement No 805186). GGK was supported by the US National Science Foundation (Grants NSF-AGS-164438, NSF-IOS-175489, and NSF-AGS-2028644).

\appendix
\section{Sensitivity analysis to the choice of $n=3$ in $c^n\Lambda$}\label{app:sensitivity}

\begin{figure}
\centering
\includegraphics[width=0.8\textwidth]{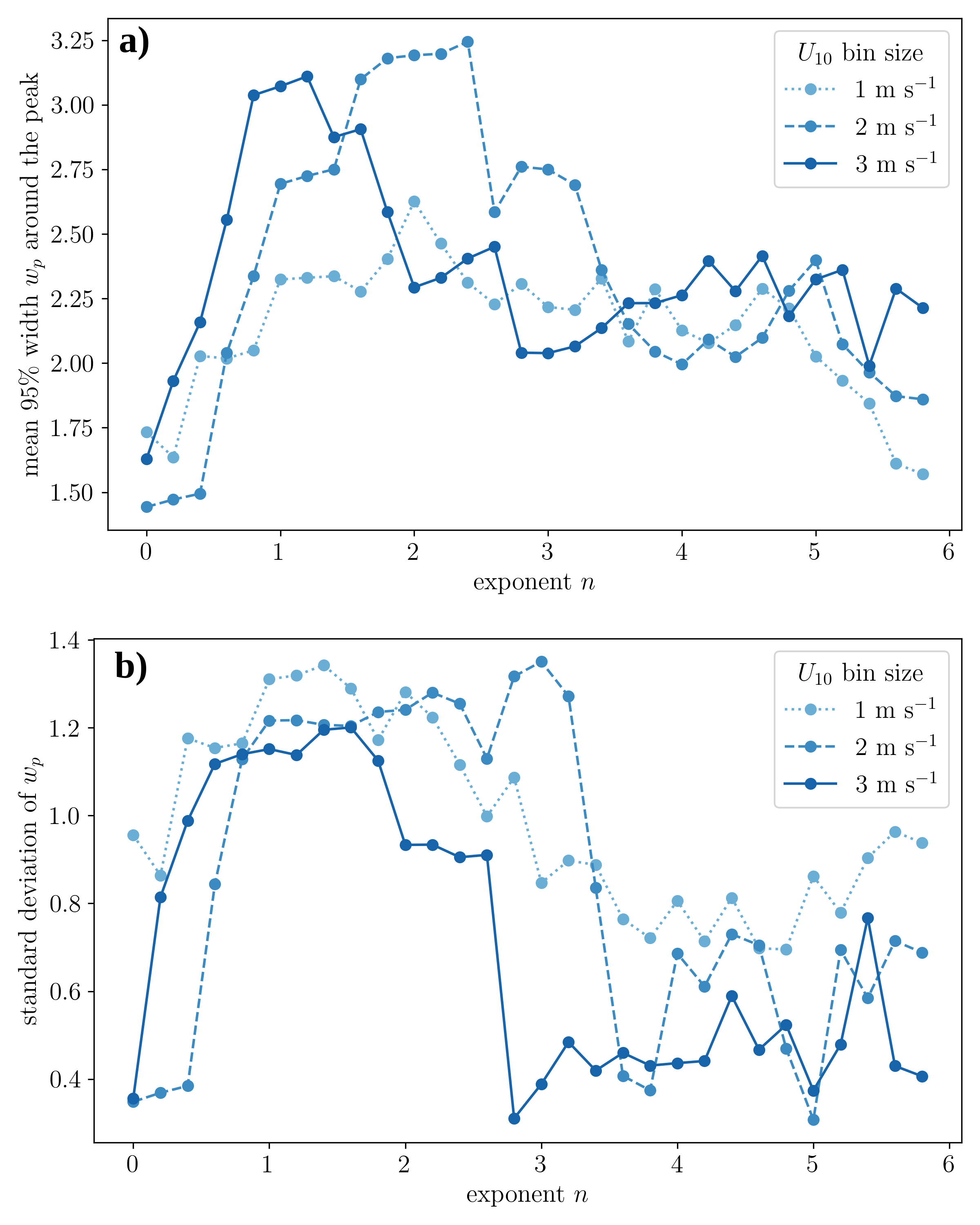} 
\caption{Width of the peak of $\partial (c^n \overline{\Lambda})/\partial U_{10}$ for different values of the exponent $n$ and different bin sizes. The width is defined as $w_p(U_{10},n) = |\log(c_u) - \log(c_d)|$, where $c_u$ and $c_d$ are respectively the smallest and largest breaker front speeds for which $\partial (c^n \overline{\Lambda}) / \partial U_{10}$ drops below 95\% of its maximal value. (a) Mean and (b) standard deviation of $w_p$ over all $U_{10}$-bins.}
\label{fig:s:width}
\end{figure}

Figure \ref{fig:lambdas}c shows that the gradient of $c^3 \overline{\Lambda}(c,U_{10})$ with respect to $U_{10}$ has a $U_{10}$-dependent peak associated with the characteristic speed $c_r(U_{10})$. In \S\ref{sec:data}, we also provided theoretical arguments justifying the choice of the exponent $n=3$ when analysing variations of $c^n \overline{\Lambda}(c, U_{10})$. To provide further support for these arguments, sensitivity tests are performed on the exponent $n$. The idea underlying the analysis of this appendix is to find the exponent for which the peak in $\partial (c^n \overline{\Lambda}) / \partial U_{10}$ is best defined.

We first quantify the width around the peak of the curves shown in figure \ref{fig:lambdas}c by computing $|\log(c_u) - \log(c_d)|$, where $c_u$ and $c_d$ are respectively the smallest and largest breaker front speeds for which $\partial (c^n \overline{\Lambda}) / \partial U_{10}$ drops below 95\% of its peak value. The natural logarithm is used since, for different $U_{10}$ values, the width of the peak is comparable only in order of magnitudes (i.e. in logarithmic coordinates, as shown in figure \ref{fig:lambdas}c). The $U_{10}$-averaged value of this width, shown in figure \ref{fig:s:width}a as a function of the exponent $n$, reveal that, besides $n= 3$, $n=0$ is also an exponent for which the peak is narrow. The variance of this width with across all measured $U_{10}$ values (figure \ref{fig:s:width}b), confirms that this statement is statistically correct.

\begin{figure}
\centering 
\includegraphics[width=0.75\textwidth]{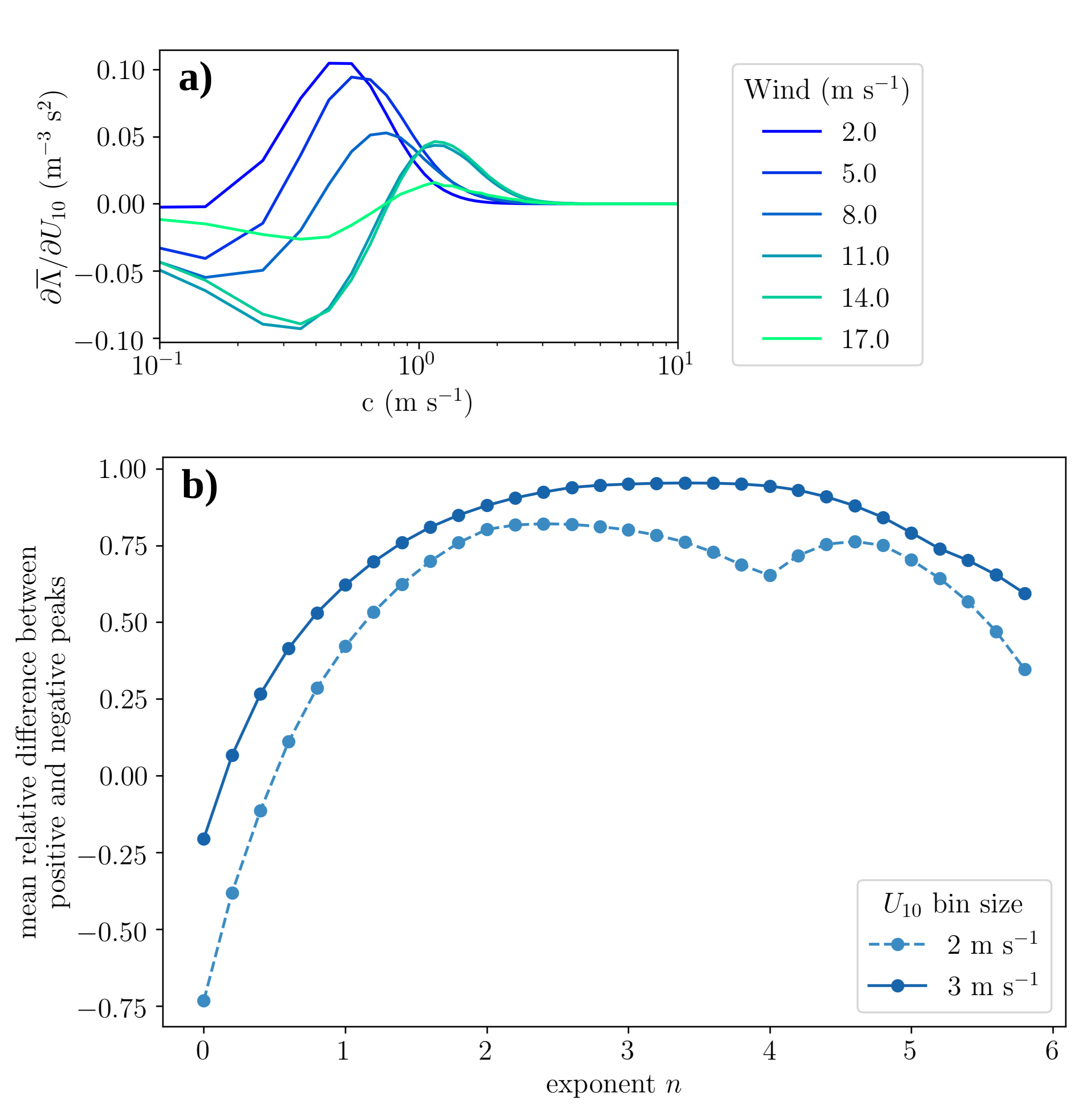} 
\caption{(a) $\partial \overline{\Lambda}/\partial U_{10}$ as a function of the breaker speed for 3 m s$^{-1}$ $U_{10}$ bins. This figure shows that, for a low value of the exponent (here $n=0$), there exists a second peak in the gradient of $c^{n}\overline{\Lambda}$ which is negative. (b) Difference between the value of the negative peak and of the positive peak, normalized by the value of the positive peak. This difference is averaged aver all $U_{10}$-bins (each corresponding to one green/blue curve in (a)), and plotted as a function of the exponent $n$. This figure shows that $n\sim 3$ is the exponent for which the negative peak is the smallest with respect to the positive peak.}
\label{fig:s:secondpeak}
\end{figure}

However, for $n < 3$, $\partial (c^n \overline{\Lambda}) / \partial U_{10}$ exhibits a second, negative peak, at lower values of $c$ (see figure \ref{fig:s:secondpeak}a for an example for $n=0$). This implies that there is no longer a unique characteristic breaker speed that can be defined from the analysis. This observation can be quantified by computing the difference between the absolute values of the positive and the negative peaks, normalized to the value of the positive peak (figure \ref{fig:s:secondpeak}b). For low $n$, the difference is negative, implying that the negative peak has a higher value than the positive peak. For $n$ close to $3$, the negative peak has a very small value with respect to the positive peak (the solid blue curve in figure \ref{fig:s:secondpeak}b has its maximum for $n~3$): this supports the choice $n=3$ and discards the choice $n = 0$.

Note that in figure \ref{fig:s:secondpeak}b we did not show results from the data binned in $U_{10}$-intervals of 1 m s$^{-1}$. This is because, for this choice of bins, the value of the positive peak is very weak for $n$ close to $0$, and hence the relative difference becomes too large to be compared to the two curves shown in Fig \ref{fig:s:secondpeak}b. Besides, binning data in $U_{10}$-intervals of 1 m s$^{-1}$ results in a low number of points per bin size.

\section{Derivation of the phenomenological turbulence model}\label{app:pheno}

The derivation of (\ref{eq:wind-shear}) is offered starting from a phenomenological model of wall-bounded turbulence \citep{gioia2010spectral,katul2011mean}. The calculations follow closely those of \citet{ayet2020impact}.

We model an idealized SBL, defined as the lowest part of the atmospheric boundary layer (adjacent to the surface) where the flow has high Reynolds number and is horizontally homogeneous and stationary with no subsidence. All averaged atmospheric quantities are invariant with respect to the streamwise and spanwise directions, and depend only on height ($z$) from the surface. For near-neutral atmospheric stability conditions ($|z/L|<0.1$, where $L$ is the Obukhov length), the Turbulence Kinetic Energy (TKE) budget equation is then a balance between mechanical production and dissipation
\begin{equation}\label{eq:tke}
u_*^2\frac{\partial U}{\partial z} = \epsilon,
\end{equation}
where $U$ is the mean wind speed (in the streamwise direction), $u_*$ is the friction velocity, and $\epsilon$ is the TKE dissipation rate.

To obtain a link between the mean wind speed and friction velocity, a closure for $\epsilon$ is needed. To this end, it was hypothesized \citep{gioia2010spectral, katul2011mean} that turbulent structures of half streamwise and vertical extension $s_e$ are attached to the surface (red circles in figures \ref{fig:stress}b,c). The turnover velocity $w_e$ of these structures was first related to $\epsilon$ using the Kolmogorov 4/5 law for the third-order velocity structure function  \citep{moninyaglom1975},
\begin{equation}\label{eq:kolmogorov}
   w_e(s_e) \propto (\epsilon s_e)^{1/3}.
\end{equation}
Then, the momentum flux $u_*^2$ through a given surface at a height $z$ (dashed lines in figures \ref{fig:stress}b,c) was expressed as a function of the vertical transport of momentum by the turbulent structures, reading
\begin{equation}\label{eq:momentum}
    u_*^2(z) \propto  w_e(s_e) [U(z+s_e) - U(z-s_e)] \propto  w_e(s_e)~ \frac{\displaystyle \partial U}{\displaystyle \partial z}~ 2s_e.
\end{equation}
The TKE dissipation was finally obtained by inserting (\ref{eq:momentum}) in (\ref{eq:kolmogorov}). Using this outcome in (\ref{eq:tke}), the scaling $w_e \propto u_*$ and (\ref{eq:wind-shear}) are easily obtained.


\section{Wave-induced motions and their link with the phenomenological model}\label{app:extend}

\begin{figure}
\centering
\includegraphics[width=\textwidth]{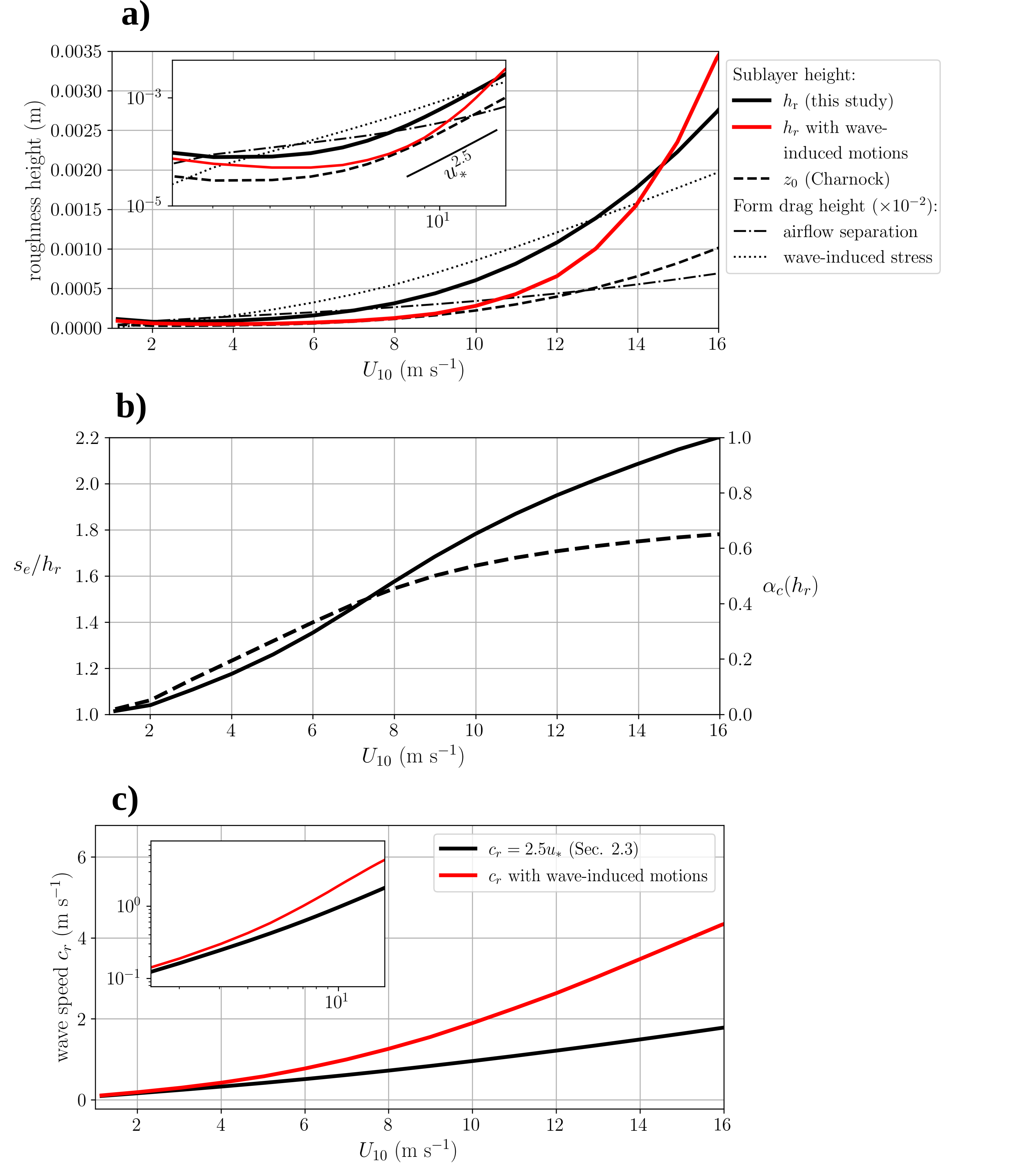}
\caption{Properties of the roughness sublayer in the presence of wave-induced motions. (a) Different sublayer heights are compared: the roughness sublayer height from this study without (solid black line) and with (solid red line) wave-induced motions, and the roughness height $z_0$ using the Charnock parameterization (dashed line). Also shown are the vertical integral length scales of airflow separation (dotted-dashed line) and wave-induced stress (dotted line) from the wind-over-wave model of \cite{Kudryavtsev2014}. (b) The ratio of the size of roughness-sublayer eddies and the height of the roughness sublayer when wave-induced motions are included in the logarithmic sublayer (solid line, left axis). This ratio and the red lines in (a) and (c) have been computed using a coupling coefficient at the top of the roughness sublayer (dashed line, right axis) from the wind-over-wave model of \cite{Kudryavtsev2014}. (c) Representative breaker speed $c_r$ with (black line) and without (red line) wave-induced motions (corresponding the roughness heights of (a)).}
\label{fig:s:height}
\end{figure}

Using open-ocean measurements of $u_*$ and $U_{10}$ and their COARE fit (solid line in figure \ref{fig:measurements}a), a wind-dependent $z_0$ can be inferred and a corresponding $h_r$ can be computed from (\ref{eq:z0-hr}). As shown in figure~\ref{fig:s:height}a (black solid line), $h_r$ ranges from $10^{-4}$ m to $10^{-2}$m. The log-log inset further reveals that both $h_r$ (black solid line) and $z_0$ (dashed line) follow a power law whose exponent is negative for winds lower than $3$ m s$^{-1}$, and tend towards  $u_*^{2.5}$ for wind speeds larger than  $8$ m s$^{-1}$. These two regimes are reminiscent of the dependence of $z_0$ with $u_*$, which is usually parameterized as \citep{charnock1955wind}
\begin{equation}
z_0  = z_0^v + \zeta_c\frac{u_*^2}{g}.
\end{equation}
where $\zeta_c$ is the Charnock coefficient, which depends on $U_{10}$ or $u_*$ \citep[see e.g.][]{edson2013exchange}. 

In wind-over-waves models \citep[e.g.][]{kukulka2007model, kukulka2008effect1, kukulka2008effect2, Kudryavtsev2014}, waves generate \emph{wave-induced motions} in the SBL due to: (i) airflow separation events associated to breaking waves \citep{reul1999air}, and (ii) the contribution of waves with low steepness that deform the overlying airflow \citep[a process called wave-induced stress in the following, see e.g.][]{Plant1982, belcher1993turbulent}. For a given wind-wave field, both mechanisms generate stresses which decay exponentially with height, and dashed-dotted and dotted lines  in figure \ref{fig:s:height}a correspond to their vertical integral length scales divided by 100. When scaled, the integral length scales of both mechanisms seems to have the same order of magnitude of $h_r$, but do not exhibit the correct power-law exponent (as shown in the log-log inset). Characterising as a whole the wind-over-wave coupling, $h_r$ does not model any of those mechanisms, but corresponds to the height of an \emph{effective roughness element} representing the bulk effect of individual wind-waves on the energy-containing eddies at the bottom of the SBL (in the roughness sublayer). 

It is however well known that above the roughness sublayer, the stresses caused by air-flow separation events and wave-induced stresses result in a deviation of the wind profile from its logarithmic form \citep[e.g.][]{miles1965note, makin1999coupled,hara2004wind, Kudryavtsev2014}. This deviation results from TKE dissipation being modified by the presence of wave-induced motions \citep[see][]{ayet2020impact}, leading to a mean wind gradient of the form 
\begin{equation}\label{eq:app:shear}
    \frac{dU}{dz}(z) = \frac{u_* [1 - \alpha_c(z)]^{3/4}}{\kappa s_e(z)}, \;\; \mathrm{for }\; z \geq h_r
\end{equation}
where $\alpha_c(z) = (u_*^2 - \overline{u'w'}(z))/u_*^2 <1$ is the coupling coefficient that accounts for attenuation of the turbulent momentum flux $\overline{u'w'}(z)$ as height decreases, due to motions being increasingly coherent with the waves (i.e. the presence of additional stresses, which induce a deviation of $\overline{u'w'}$ from its value $u_*^2$ at the top of the SBL). 

In the roughness sublayer, as mentioned in \S\ref{sec:pheno}, it is assumed that turbulent motions are indistinguishable from wave-induced motions, and hence 
\begin{equation}
    \frac{dU}{dz}(z) = \frac{u_*}{\kappa s_e}, \;\; \mathrm{for }\; z \leq h_r.
\end{equation}

The deviation of the mean wind profile affects the wind speed at the bottom of logarithmic layer and hence the properties of the roughness sublayer. Indeed, the condition of continuity for the mean wind shear at the top of the roughness sublayer leads to the following relation
\begin{equation}\label{eq:sh:anisotropic}
    s_e = h_r[1-\alpha_c(h_r)]^{-3/4}.
\end{equation}
As discussed in \S\ref{sec:pheno}, this interesting relation reveals that the properties of roughness-sublayer energy-containing eddies are coupled with wave-induced motions aloft.

In general, the dependence of $\alpha_c$ with height and wind speed is non trivial and analytical expressions of $h_r$ dcannot be derived. In figure~\ref{fig:s:height}a, red line, we show  $h_r$ using a wind-dependent $\alpha_c(h_r,U_{10})$ from the wind-over-waves model of \citet{Kudryavtsev2014} (dashed line in figure \ref{fig:s:height}b). In this model, $\alpha_c$ includes both airflow separation events and wave-induced stress to yield a momentum flux and short wind-wave spectrum consistent with measurements and the COARE parameterization. Figure~\ref{fig:s:height}a reveals that the power-law dependence of $h_r$, discussed above, is affected by the presence of wave-induced motions in the logarithmic sublayer (see the inset of figure \ref{fig:s:height}a). However, the order of magnitude of $h_r$ is similar to that when wave-induced motions are neglected (compare black and red lines). When translated in terms of critical wind speed $c_r = U(h_r)$, the disagreement is similar (figure \ref{fig:s:height}c). Finally, in figure~\ref{fig:s:height}b, we further show the resulting ratio between $s_e$ and $h_r$ (solid line), discussed in \S\ref{sec:pheno}.

For completeness, a simplified case for which the roughness sublayer height can be computed explicitly is illustrated. To that end, a simplified height dependence of wave-induced stress is assumed following \citet{makin1995drag}, and is given as 
\begin{equation}
    \alpha_c(z) = \alpha_c^0 \mathcal{H}(h_w - z),
\end{equation}
where $\mathcal{H}$ is the Heaviside step function, and $h_w$ is the height at which the effect of waves on momentum flux becomes negligible. This expression assumes that the coupling coefficient is constant and equal to $\alpha_c^0$ below a height $h_w$, and zero above.

Equation (\ref{eq:app:shear}) reveals that wave-coherent motions induce a deviation from the logarithmic profile. Nevertheless, by extrapolating a logarithmic profile, a roughness height $z_0$ can still be defined from the wind at an arbitrary height $z$, $U(z) = (u_*/\kappa)\log(z/z_0)$.  Upon using this expression, the simplified height dependence of $\alpha_c$, and integrating (\ref{eq:app:shear}) from an arbitrary height $z$ down to $h_r$, the mean wind speed at the roughness sublayer height reads
\begin{equation}
    U(h_r) = \frac{u_*}{\kappa}\log\left(\frac{h_r^{(1-\alpha^0_c)^{3/4}}}{z_0} h_w^{1-(1-\alpha^0_c)^{3/4}} \right)
\end{equation}

By further using (\ref{eq:sh:anisotropic}), the mean wind speed and the roughness sublayer height can be expressed as
\refstepcounter{equation}
$$
U(h_r) = (u_*/\kappa)(1 - \alpha_c(h_r))^{-3/4}
    , \quad 
h_r = z_0^{(1-\alpha^0_c)^{-3/4}}h_w^{1 - (1-\alpha^0_c)^{-3/4}}e^{(1 - \alpha_c^0)^{-3/4}}.
  \eqno{(\theequation{\mathit{a},\mathit{b}})}
$$
The above expressions show that as $\alpha_c$ increases, the roughness sublayer height increases from $z_0\exp(1)$, and so does the mean wind speed on top of the roughness sublayer.  The limit $\alpha_c \to 1$ is out of the range of the model, since the mean wind shear can no longer be continuous at $h_r$ in this regime.

\bibliographystyle{jfm}
\bibliography{scale}

\end{document}